# De-risking renewable energy investments:
## Assessing contract design and project finance using operational wind park data

Jorge Sánchez Canales [a,*], Lion Hirth [a,b]

[a] *Centre for Sustainability, Hertie School, Friedrichstraße 180, 10117 Berlin, Germany*
[b]*Neon Neue Energieökonomik GmbH, Schönleinstraße 31, 10967 Berlin, Germany*
* Corresponding author: j.sanchez-canales@hertie-school.org

**Abstract**

Investment in renewable electricity generation is highly capital intensive and therefore strongly dependent on financing conditions. In Europe, much of this investment has occurred under public support schemes that resemble long-term public contracts such as feed-in tariffs (FiTs) and contracts-for-differences (CfDs). These contracts not only subsidize renewable generation but also stabilize project cash flows by reducing exposure to electricity price volatility, thereby improving debt capacity and lowering financing costs. At the same time, they may distort operational and investment incentives by weakening exposure to wholesale market price signals. This paper studies how alternative public contract designs reduce revenue risk and how this translates into financing outcomes. Using a novel dataset of hourly turbine-level generation covering 63 German onshore wind parks over the period 2013–2024, we simulate project cash flows under two-sided CfDs, one-sided CfDs, and financial CfDs. We then evaluate their implications for cash-flow volatility, debt capacity, and the levelized cost of electricity using a project finance model based on a conservative debt-service coverage ratio (DSCR) constraint. We find that financial CfDs provide hedging performance comparable to conventional two-sided CfDs. The results suggest that the commonly assumed trade-off between revenue stabilization and efficient market integration is not inherent but depends on contract design. More broadly, public contracts can substitute missing long-term hedging markets. These results have direct policy implications for the design of renewable energy support schemes.

# 1. Introduction

The vast majority of global power generation investment today goes into wind and solar energy. These technologies are characterized by high upfront capital costs and low operating costs. Consequently, small changes in the cost of capital can substantially affect the levelized cost of electricity (LCOE) and project viability. Yet low-cost financing requires stable and predictable revenues, which is at odds with the volatility of wholesale electricity prices and stochastic renewable generation.

To address this problem, governments across Europe have relied on policies that can be described as long-term public contracts such as feed-in tariffs (FiTs) and contracts for differences (CfDs). While commonly discussed as subsidy instruments, these contracts also perform a second and equally important function: they hedge revenue risk. By reducing exposure to future electricity price volatility, long-term contracts improve the bankability of renewable projects and enable higher leverage ratios in their financial structures.

This financial mechanism is particularly important because long-term electricity price risk cannot be efficiently hedged in private markets. Financial electricity products are typically traded only up to 5 years into the future, whereas renewable project finance must plan for investment horizons of 25 to 35 years. Public support contracts therefore substitute for missing long-duration hedging markets by reallocating price risk from investors to the public sector.

However, revenue stabilization may come at the cost of distorted operational and investment incentives. Under fixed-price support schemes such as FiTs or simple two-sided CfDs, generators may continue producing even during periods of negative wholesale prices because revenues are disconnected from market conditions. Similar distortions emerge in investment decisions. For example, solar installations optimized solely for production volume (i.e., facing South in the northern hemisphere) may receive the same per-unit remuneration as installations better aligned with electricity demand (i.e., facing East or West). By removing price signals, highly protective contracts can therefore create welfare losses and inefficient capital allocation despite improving financing conditions.

Different contract structures provide different degrees of downside protection, upside retention, and exposure to spot-market prices, implying heterogeneous effects on both financing conditions and market incentives. Recent literature therefore frames renewable contract design as a trade-off between revenue stabilization and efficient market integration (Fabra & Llobet, 2025; Favre & Roques, 2023; Johanndeiter et al., 2025; Leblanc, 2023). Among others, Schlecht et al. (2024) have proposed a novel "financial CfD" which aims at resolving the trade-off by keeping the exposure to market prices, but they did not provide assessment of its de-risking performance.

While a growing literature studies risk mitigation under CfDs (Alcorta et al., 2024; Chebotareva et al., 2020; Đukan et al., 2025; Fabra & Llobet, 2025; Falconett & Nagasaka, 2010; Johanndeiter et al., 2025), these studies are mostly theoretical or rely on simulated data, and empirical evidence remains scarce. An important contribution is the paper by Đukan et al. (2025) which uses synthetic data and financial modeling to show how two-sided CfDs reduce cash-flow volatility and lower the cost of capital.

In this article, we make three contributions to the empirical literature on renewable energy contracts. First, we provide a comprehensive comparison of major contract types, specifically two-sided CfDs, one-sided CfDs, and financial CfDs. These contracts differ with respect to their incentives for market integration and the cash flows they generate, but in this paper we focus only on the latter. Second, we build on Đukan et al. (2025) and develop a corporate finance model to assess how contract design affects the distribution of revenues, capital structure, and the cost of capital, allowing us to quantify implications for LCOE. Third, we employ novel empirical data, rather than relying on weather simulations. We use a unique dataset from a major German operator covering hourly electricity production (realized and potential) from more than 60 onshore wind parks over the period 2013–2024. This is important because, as practitioners often argue, real-world production restrictions due to wildlife protection, shadow and noise impact on nearby human settlements, localized grid bottlenecks or turbine downtime render simulated data unreliable. These frictions are included in our revenue analysis, ensuring that the comparison of contracts reflects the actual risk profiles faced by real-world wind projects.

Our findings show that optimally designed financial CfDs outperform current one-sided designs in reducing cash-flow risk and can, depending on the specifications, outperform the standard two-sided CfD design as well. This suggests that the commonly assumed trade-off between revenue stabilization and market integration can be resolved using smart contract design.

# 2. Literature review

We first review the literature linking cash flow risk and renewable energy financing, then studies on CfD design, in particular empirical studies, before placing the present paper in the landscape of exiting literature.

## *2.1. Cash-flow risk and renewable energy projects*

High cash-flow volatility not only reduces the amount of internal funds available to firms but also reduces the opportunities to access external financing, for instance increasing borrowing costs and the likelihood of a firm going into financial distress (Li et al., 2023). Several empirical finance studies support this link between volatility and credit spreads (e.g., Minton et al., 2002; Minton & Schrand, 1999; Molina Manzano, 2005; Tang & Yan, 2010), often using the coefficient of variation (CoV) of annual revenues as a normalized measure of cash-flow volatility. We follow this body of literature and employ CoV as the metric to measure cash-flow risk in the first part of our analysis (Section 3.2). At the same time, project finance practice and credit risk models emphasize downside risk rather than symmetric variability, since default and covenant breaches are triggered by low-cash-flow states rather than high volatility per se (Augustin et al., 2025). This motivates our focus on the minimum annual cash flows in the financial model below (Section 3.3).

Wind energy projects are capital-intensive, which makes them highly sensitive to fluctuations in cash flows. There are three specific sources of uncertainty affecting the stability of cash flows in wind projects: market (price) risk, weather-related (volume) risk, and grid-related curtailment (volume) risk (Gatzert & Kosub, 2016). Among these, weather-related uncertainty has attracted the most attention in the literature, given its stochastic nature and direct impact on generation levels (Collins et al., 2018; Staffell & Pfenninger, 2018). Public support schemes, however, primarily target price and curtailment risk, leaving volume risk largely unmanaged.

Regarding renewable energy investment projects, a key contribution is Đukan et al. (2025), who model an off-shore wind farm to show the impact of revenue stability in bringing down the cost of capital for investors. Their analysis links revenue profiles to financing terms but relies on simulated generation and prices rather than observed project data.

### 2.2. Contract for difference design

A recent strand of analytical work has been exploring the optimal design of public contracts for differences for renewable energy. A key insight from this literature is the risk-efficiency trade-off: full price insulation reduces risk but weakens incentives to respond to market conditions, while full exposure to market prices maximizes allocative efficiency but increases financing costs due to higher risk premia (Favre & Roques, 2023). The major innovation was the idea of using an exogenous capacity payment combined with retained earnings from market operation, since this combination leads to lower risk while keeping system-aligned incentives for dispatch and investment (Barquín Gil et al., 2017; Huntington et al., 2017; Newbery, 2023). Specifically, Fabra and Llobet (2025) argue that using some two-way fixed capacity payments to complement market revenues can simultaneously address the problems of adverse selection and moral hazard:[1] they adjust remuneration without introducing revenue risk, while at the same time can be used to limit excessive rents in publicly subsidized investments.

There have been several recent proposals of CfD designs that use these elements – exogenous capacity remuneration and reference profiles. Examples include Newbery's (2023) "yardstick CfD", the "capability CfD" promoted by the Elia Group (ENTSO-E, 2024), or the long-term framework agreed by the European Commission for the Czech nuclear expansion, which combines a regulated capacity payment with market-based revenues (European Commission, 2024). Among these, in this article we focus on the "financial CfD" proposed by Schlecht et al. (2024). This last design similarly decouples payments from actual production by tying the net market income to an exogenous benchmark generation profile to avoid distortions in plant operations. On top of that, the contract uses a lump sum capacity-based payment to further smooth the cash-flow profile. This contract design is consistent with the recommendation by Fabra and Llobet (2025) that the optimal contract combines per-unit and per-capacity payments, and the usage of external reference profiles. In the article, the authors highlight that the net risk implications of such contracts are ultimately an empirical question – a question this paper addresses by using real-world generation data to evaluate the effectiveness of financial CfDs in reducing revenue volatility.

[1] In this context, adverse selection refers to the government's inability to differentiate between the quality of assets when investors bid in auctions. This means that the government cannot assign different strike prices to different technologies. Moral hazard occurs when, after signing the contract, the government cannot influence the effort that a generator will put into producing electricity during periods of high prices.

In this article, we focus on three representative designs: a classic two-sided CfD, widely used in European renewable auctions since the mid-2010s; a one-sided CfD, where public support only tops up revenues when prices fall below the strike, which has been used in Germany since 2017; and a more recent "financial CfD" variant that adds capacity-style payments to market earnings. Section 3.1 contains a detailed description of the different contract designs we model.

### *2.3. Quantitative studies*

Despite the existence of analytical models, empirical work remains limited, particularly when it comes to real-world data and several contract designs. Partial exceptions are Favre and Roques (2023), Johanndeiter et al. (2025) and Đukan et al. (2025).

Favre and Roques (2023) analyze how alternative CfD designs reshape the trade-off between investment incentives and risk allocation. Focusing on "adapted" CfDs –variants that decouple contractual volumes from actual production–, their study uses a cross-European power system model with Monte Carlo simulations to compare revenue profiles for nuclear, onshore wind, and solar under different contract designs. The authors conclude that policymakers face a fundamental design choice: stronger incentives require greater investor exposure to market risk. We follow their approach by comparing multiple contract designs.

Similarly, Johanndeiter et al. (2025) simulate wind park revenues under various CfD designs, including two-sided, one-sided, and financial, but they take a wider perspective of total system costs. While they try to assess how contract design affects strike prices, investor revenues, and electricity market risks across different bidding zones, their major contribution is to integrate CfD design into a system-level modeling framework. Their simulations, however, are based on a limited set of modeled scenarios (for instance, they use a single weather year across all simulations), which constrains the extent to which they can speak of empirical cash-flow risk at the project level.

Đukan et al. (2025) is particularly close to our study, since they also focus on the risk implications of CfDs. They develop a project finance model for a single offshore wind farm in the German North Sea and compare a merchant baseline to a classic two-sided CfD, combining stochastic power price and feed-in simulations with cash-flow-based debt sizing. They show that a two-sided CfD substantially reduces revenue volatility, lowers the probability of financial distress, increases debt capacity and lowers WACC, thereby quantifying the financing benefits of revenue stabilization for a single project. Our study takes inspiration from their financial model, and we develop our own model to link revenue stability to financing conditions.

### 2.4. Gap and contribution

While theoretical work has advanced optimal CfD design and simulations have begun to quantify risk-incentive trade-offs, there is still need for empirical evidence on realized cash-flow volatility under multiple designs and its translation into project-financing metrics relevant for investors and policy makers. Our study fills this gap, using a unique dataset of hourly realized and potential generation from 268 turbines across 63 onshore wind parks in Germany (2005–2024), which captures real-world frictions like curtailment, downtime, and site heterogeneity that simulations often overlook.

Building on the theoretical insights and the previous empirical papers presented above, our paper contributes to the empirical literature on CfDs by combining for the first time three key features:

- Multiple contract designs: Like Favre & Roques (2023) and Johanndeiter et al. (2025), we compare two-sided, one-sided, and financial CfDs.
- A corporate finance model: We use a stylized financial model to translate the insight of cash-flow stability, and particularly of downside risk, into relevant financial indicators such as project-level WACC and LCOE reductions.
- Real-world data: Unlike synthetic simulations in all prior studies, we employ historical production data. Real-world production is shaped by a range of constraints that are often unobserved or difficult to model: turbine maintenance schedules, curtailment due to environmental permitting (e.g., bird or bat protection), noise or light emission limits, and local grid restrictions. These factors introduce deviations from the theoretical production potential of a site and are highly relevant for understanding the risks investors face. Moreover, our dataset allows us to capture a key source of heterogeneity behind investments, not only of wind and price years as is common in simulation studies (Đukan et al., 2025; Favre & Roques, 2023; Johanndeiter et al., 2025), but also of the quality of sites and turbine technologies. By accounting for operational frictions and turbine-level heterogeneity, we provide a more robust study of relative contract performance compared to previous studies. Each year in the dataset can be interpreted a distinct realization of a "state of the world," characterized by a unique combination of weather, market, and operational conditions. This way, we seek to bridge contract design (Fabra & Llobet, 2025; Newbery, 2023; Schlecht et al., 2024) and practitioner concerns about revenue stability in actual wind investments.

No prior study combines these three features simultaneously, so we provide the first empirical benchmark tying novel CfD designs to observed cash-flow distributions for a large fleet of operational wind projects.

# 3. Method

In this article we measure how different CfD designs hedge cash-flow risk for wind park investors and how this translates into financing conditions. First, we present the set of contract designs we use to simulate hourly and annual revenue streams for each wind park in our sample using historical production and price data. Second, we summarize the resulting cash-flow volatility under each contract with the coefficient of variation of annual revenues. Third, we develop a stylized project finance model with endogenous debt sizing that maps the simulated revenue distributions into leverage, the cost of capital (WACC), and the implied levelized cost of electricity (LCOE).

## *3.1. Contract designs*

We assess ten different contract specifications in total, corresponding to three groups: two-sided CfDs, one-sided CfDs, and financial CfDs. As a benchmark, we use the merchant case (no long-term contract).

### Merchant case

As "merchant case" $M$ we describe the situation that a generator has no long-term contract and receives revenue only from selling output to the electricity spot market. The revenue $R$ of a generator $i$ for an hour $t$ is defined as the product of the generation $q_{i,t}$ and the electricity spot price $p_t$. For simplicity, we assume that for each hour there is a unique spot price, which can be interpreted as generators selling their production only on the day-ahead auction of the wholesale market. In other words, we ignore intraday markets and imbalance settlement. The revenue over the course of year $y$ is given by

$$R_{i,y}^{M} = \sum_{t=1}^{t=T} q_{i,t} p_t \,, \tag{1}$$

where the set $t = \{1, T\}$ refers to all hours in year $y$.

### Two-sided CfD

The two-sided CfD, which we denote $2CfD$ in equations, is a contract design that was introduced in the UK electricity market reform of 2014. The idea of a CfD is to top up the market income by comparing a reference price to a pre-agreed strike price. The government pays the generator when the reference price falls below the strike price, and the generator pays the government when the reference price exceeds the strike price. This effectively guarantees a per-unit sale price equal to the strike.

We decompose hourly revenue into market income and CfD payments. For a given hour $t$, the total revenue of a generator $i$ is defined as the product of generation $q_{i,t}$ and the sum of the realized spot price ($p_t$) and the CfD payment, which depends on the difference between the strike price ($S$) and the reference price ($p_t^{ref}$). The total annual revenue is

$$R_{i,y}^{2CfD} = \sum_{t=1}^{t=T} q_{i,t}p_t + q_{i,t}(S - p_t^{ref}). \quad (2)$$

Under a classical "UK-style" two-sided CfD, the reference price equals the hourly spot price at which generators sell their electricity in the spot market ($p_t^{ref} = p_t$), such that the CfD payment simply offsets deviations of $p_t$ from the strike. This case constitutes our first specification. In this case there is only one source of variation, the *volume risk* of production $q_{i,t}$. The price risk disappears since the strike price remains the same for the duration of the contract. This contract leads to the so-called "produce-and-forget" incentives: the investor has only an incentive to maximize the volume of production $q_{i,t}$ regardless of the economic value of the energy produced (which depends on the time when the electricity was produced).

Because this variant of the contract is rarely considered by policymakers nowadays, we model a second specification using an annual reference price. The reference price is the annual capture price (market value) of onshore wind, defined as the volume-weighted average price of onshore-wind-generated electricity across all wind parks in the country:

$$\bar{V}_y^W = \frac{\sum_1^T (Q_t^w p_t)}{\sum_1^T Q_t^w}, \quad (3)$$

where $Q_t^w$ refers to the market wide generation of onshore wind power. When a reference price other than the hourly spot market is used, different incentives emerge. Because the reference is exogenous to a single plant but correlated with its output, it provides an imperfect but good hedge against price risk while keeping some of the wholesale price incentives alive. In hours when the spot price falls significantly below zero and the strike is not high enough to justify production ($S + p_t - p_t^{ref} \leq 0$), we assume that generators curtail all production ($q_{i,t} = 0$) to avoid losses.

Our third and last specification of the two-sided CfD suspend payments from governments to generators whenever spot prices turn negative. This incentivizes generators to always curtail their production at negative prices ($p_t \leq 0$). This is the type of CfD the European Commission currently prefers (Butorac, 2026).

In sum, we assess three variants of the two-sided CfD: an hourly reference price, an annual reference price, and annual reference in addition to suspension during negative prices. In this order, incentives improve but hedging quality is suspected to deteriorate.

## One-sided CfD

Under a one-sided CfD ($1CfD$), also known as the "market premium model", which has been used in Germany since 2012 but also in The Netherlands and in Italy, generators receive payments from the government but are not required to make reverse payments.

We use the same decomposition into market income and CfD payments as above. The overall annual revenue a generator producing electricity under a one-sided CfD is

$$R_{i,y}^{1CfD} = \sum_{t=1}^{t=T} q_{i,t}p_t + q_{i,t} \cdot \max\left(S - p_t^{ref}, 0\right). \tag{4}$$

Under this contract, the price risk is only partially mitigated. In the extreme case where $p_t^{\mathrm{ref}} \geq S$ for all $t$, the contract is identical to that of having no support, whereas in the opposite extreme case where $p_t^{\mathrm{ref}} \leq \mathrm{S}$ for all $t$, the contract is identical to that of a two-sided CfD.

As with the two-sided CfD, we consider three specifications:

(i) using the hourly spot price $p_t$ as the reference price $p_t^{\mathrm{ref}}$;
(ii) using the annual capture price of wind $\bar{V}_y^W$ (defined above in equation 3) as the reference price $p_t^{\mathrm{ref}}$; and
(iii) using the capture price of wind $\bar{V}_y^W$ while suspending payments when spot prices are negative.

As above, we assume generators always dispatch according to economic incentives, i.e. curtail if total revenue is negative.

## Financial CfD

The third major design we consider is the financial CfD, as proposed by Schlecht et al. (2024). This contract consists of two separate hourly payments between generator and government: a fixed (lump-sum) payment from the government to the generator $F$ in exchange for a variable (floating) payment from the generator to the government that reflects market conditions $V_t$. Effectively, the contract is a fixed-for-floating swap.

Unlike conventional CfDs, this contract is not tied to asset outputs, hence it has been described as "injection-independent" (as opposed to "production-based"). For a generator with installed capacity $K_i$ (in MW), the contracted capacity $k_i$ can be defined as:

$$k_i = \alpha \times K_i, \tag{5}$$

Where $\alpha > 0$ is the contract factor.

At each hour $t$, the three revenue components are:

- The market income of the generator:
$$M_{i,t} = q_{i,t}p_t, \tag{6}$$
- The lump-sum payment from government to generator:
$$F = k_i A, \tag{7}$$

where $A$ is the payment in €/MW agreed in the contract. It is somewhat analogous to the strike price $S$ in production-based CfDs.

- The payment from generator to government:

$$V_t = k_i g_t^{ref} p_t, \tag{8}$$

where $g_t^{ref}$ is the reference generation profile (a normalized capacity factor), in MWh/MW. It is zero when the price is negative, hence $V_t$ payments are suspended.

The annual revenue of a generator under a financial CfD is

$$R_{i,y}^{f-CfD} = \sum_{t=1}^{t=T} (M_{i,t} + F - V_t) = \sum_{t=1}^{t=T} (q_{i,t} p_t + k_i A - k_i g_t^{ref} p_t). \tag{9}$$

As contract factor $\alpha$ and contracted capacity $k$ approach 0, such that $F = V_t = 0$, the revenue and risk profile of this contract converge to the merchant case. Conversely, for a generator whose production follows exactly the adjusted reference generation $q_{i,t} = k_i g_t^{ref}$, we have $M_{i,t} = V_t$ and the hourly income equals the fixed payment $F$, so revenues become completely independent of price and volume fluctuations. This highlights that the remaining source of risk under a financial CfD stems from deviations of actual production from the reference profile ("basis risk").[2]

Two parameters need to be specified to define the contract: the contract factor $\alpha$ and the reference profile $g_t^{ref}$ ($A$ is, like $S$, determined by an auction). We use the national hourly capacity factor of onshore wind as the reference profile:

$$g_t^{ref} = \frac{Q_t^w}{K_t^w} \tag{10}$$

where $K_t^w$ is the total installed capacity for onshore wind in Germany at time $t$.

By varying the way in which we set the contract factor, we model four specifications of the financial CfD. In contrast to our specifications for two- and one-sided CfDs, which affect both dispatch incentives and cash-flow risk, the financial CfD specifications only change the level of basis risk and do not affect dispatch: under this contract, generators always have an incentive to stop producing at negative prices. We therefore focus on how policymakers might choose the contract factor $\alpha_i$ to shape risk and support.

In our results, we consider the following four specifications of the financial CfD:

[2] As with all contracts that entail a payment from generator to government, this design does not imply the necessity that it is a subsidy. for a given value of $A = \frac{1}{T}\sum_t^T (g_t^{ref} P_t)$ the generator would perceive, on average, only its own market income $M_{i,t}$, making this contract revenue neutral while still smoothing (or not) the cash flow.

(i) Baseline ($\alpha_i = 1$): All parks use their full installed capacity as contracted capacity.

(ii) Fleet-optimal ($\alpha_i = \alpha^* = 0.55$): All parks share a common contract factor chosen to minimize average investment cost (equivalently, average downside risk) across our sample; the calculation of $\alpha^*$ is described in Section 3.3.

(iii) Technology-adjusted: Contract factors are park-specific and proportional to expected production levels. We approximate technology differences by each park's average annual capacity factor over the first three years after installation and scale these relative to the national capacity factor, obtaining a park-specific factor $\alpha_i^{tech}$. This specification aims to align fixed payments and claw-backs with expected production, thereby mitigating adverse selection concerns.

(iv) Park optimal ($\alpha_i = \alpha_i^*$): Finally, we allow for park-specific optimal contracts, where $\alpha_i$ is chosen specifically to minimize the LCOE for each project. The optimization procedure is detailed in Section 3.3.

### *3.2. Measuring cash-flow risk*

We assess the cash-flow risk that investors face under each contract design. Cash-flow risk is a standard concept in the corporate finance and financial distress literature. As discussed in Section 2.1, many studies measure it using the standard deviation of earnings normalized by average sales, that is, the coefficient of variation (CoV) of cash flow. We follow this literature and use the CoV of annual revenues across the sample period as a normalized measure of cash-flow risk for each wind park and contract.

The coefficient of variation (CV) is defined as

$$CV_i = \frac{\sigma\left(R_{i,y}^c\right)}{\mu\left(R_{i,y}^c\right)}, \tag{11}$$

where $R_{i,y}^c$ is the annual revenue in year $y$ under contract $c$ expressed in €/MW of installed capacity, $\sigma(\cdot)$ denotes the standard deviation, and $\mu(\cdot)$ is the mean of all years. Normalizing by average revenue allows us to compare risk across parks with different sizes and production profiles.

We simulate hourly revenues and aggregate them to the annual level before computing CoVs. Aggregation to annual revenues removes intra-year seasonality and focuses on the time horizon relevant for meeting debt service and other yearly obligations (Alcorta et al., 2024). We interpret each observed year as a distinct and equally likely state of the world, so the CoV summarizes the dispersion of revenues across these states. This provides a simple, comparable measure of overall cash-flow risk, which we complement with a project finance model focused on downside risk in the next subsection.

### 3.3. Financial model for downside risk

While the CoV is a measure of risk, policy makers and investors ultimately care about how such risk affects debt-to-equity ratios, the cost of capital, and LCOE. In project finance practice, lenders size debt based on low-percentile cash flows or minimum debt service coverage ratios (DSCRs), rather than on average volatility alone. In this spirit, we develop a stylized financial model that maps the simulated annual revenue distribution under each contract into feasible capital structures and implied financing costs (WACC). The purpose is to formalize a simple and transparent mechanism through which cash-flow risk affects those indicators.

Consider a representative wind project with a lifetime of $Y = 30$ years. Investment costs $C^{fix}$ are incurred upfront, while annual operating and maintenance costs $C^{om}$ are constant over time. Annual revenues $R_{i,y}$ are uncertain and depend on the spot market and the long-term contract, as described in Section 3.1, with each observed year interpreted as an equally likely and independent state of the world within the context of this project finance model.

Project cash flows in each year are allocated in the following order:

1. Operating and maintenance costs, assumed to be fixed and paid with certainty;
2. debt service, consisting of a constant annual payment; and
3. equity payouts, the residual cash flow.

Projects are financed through debt and equity with fixed and exogenous returns of 1.15% and 10%, respectively (see Table 1). We assume no reserve accounts and no refinancing. Debt is modeled as a fixed annuity loan with maturity equal to the project lifetime. Let $D$ denote total debt, and $r_d$ the cost of debt. Annual debt service is given by:

$$DS = D \cdot \frac{r_d (1 + r_d)^Y}{(1 + r_d)^Y - 1}. \tag{12}$$

| Parameter | Symbol | Value | Source |
|---|---|---|---|
| **Investment cost** | $C^{fix}$ | 1,500 €/kW | (IRENA, 2024) |
| **O&M cost** | $C^{OM}$ | 50 €/kWa | (IRENA, 2024) |
| **Cost of debt** | $r_d$ | 1.15 % | (Đukan et al., 2025) |
| **Cost of equity** | $r_e$ | 10.0 % | (Đukan et al., 2025) |
| **Lifetime** | $Y$ | 30 years | Own assumption |

Table 1. Cost and lifetime assumptions used in the financial model. O&M costs include the land rents that investors pay every year, which represent a very substantial share of the figure. Technical O&M are significantly lower.

The solvency condition we impose is that debt service must be fully covered in every simulated year, i.e.,

$$R_{i,y}^{c} - C^{om} \geq DS \quad \forall y. \tag{13}$$

This corresponds to a DSCR of at least one in all states. While this assumption is stricter than some standard project finance benchmarks, it allows us to directly link revenue volatility to maximum feasible leverage in a transparent way. For a given contract design $c$ and wind park $i$, this condition defines the maximum sustainable debt level $D_{i,c}^{max}$. Intuitively, contracts that raise the minimum annual cash flow permit higher leverage, which reduces WACC.

The remaining investment cost is financed with equity. Let total investment cost be $I_i = C^{fix} * K_i$. Equity is then $E_{i,c} = I_i - D_{i,c}^{max}$. Equity holders receive the residual cash flow after debt service and O&M. Their participation constraint requires the expected present value of dividends to cover the equity investment,

$$\sum_{y=1}^{Y} \frac{\mathbb{E}[\Pi_{i,c,y}]}{(1+r_e)^y} \geq E_{i,c}. \tag{14}$$

where $\Pi_{i,c,y}$ denotes annual dividends and $r_e$ is the required return on equity. We assume equity holders bear the residual risk, so we evaluate their participation in expectation, in contrast to debt which is constrained by the worst realization.

In equilibrium, competitive auctions for support contracts imply that projects bid such that they just break even. For each wind park and contract design, we jointly determine:

- the maximum feasible debt level $D_{i,c}^{max}$,
- the implied equity requirement $E_{i,c}$, and
- and the support level as defined by contract parameters (strike price or fixed payment)

such that both the debt solvency condition and the equity participation constraint are just binding. In other words, competition for contracts ensures that equity investors do not earn excessive profits. This yields an implied LCOE for each contract and project, reflecting both the distribution of annual cash flows from the contract and the endogenous financing structure. Contracts that reduce downside cash-flow risk allow projects to carry more debt, substituting cheaper debt for expensive equity, thus lower the weighted average cost of capital, and hence have a lower LCOE.

For financial CfDs, we apply the same procedure but, as discussed in Section 3.1, in some specifications we also optimize over the contract factor $\alpha_i$ to study how different parameter choices affect leverage and LCOE.

### 3.4. Data

A key strength of this study is the use of real production data from operational turbines, rather than simulated or synthetic generation profiles. Real-world production is shaped by a range of constraints that are often unobserved or difficult to model. Hence, using realized in-feed allows us to observe these constraints directly and produce more realistic assessments of revenue volatility. Our dataset also allows us to capture a key source of heterogeneity behind investments, not only of wind and price years, but also of the quality of sites and turbine technologies.

The core dataset consists of turbine-level production records for 381 wind turbines, aggregated into 93 wind parks, located in Germany. These turbines vary in size and age, with an average rated power of 2.1 MW. The oldest asset in the dataset was commissioned in December 1995, and the newest in May 2024. To ensure a sufficiently long and balanced observation period, we exclude parks entering operation after 2014 and remove parks with missing or persistently very low production observations. This way, the sample covers a ten-year period (2014–2023) during which all wind parks are continuously observed. As a result, the sample period includes major market events such as Covid-19 (low electricity prices) the energy crisis of 2022–2023 (high prices), as well as a high diversity of weather years as well as potentially unobserved variation (e.g., technical outages). Additionally, we removed parks with missing or persistently very low (potential) production profiles (with an hourly average of less than 0.06 MWh of production for every MW of rated power). The resulting estimation sample contains 268 turbines across 63 wind parks with a combined installed capacity of approximately 510 MW. The dataset and its characteristics are detailed further in Appendix A.

Our simulations use the additional time series data to calculate cash flows under each contract:

- **Hourly electricity prices** ($P_t$) from the ENTSO-E Transparency Platform.
- **Feed-in profiles** ($X_t^w$) of potential generation from Netztransparenz to estimate the national hourly production profile for financial CfDs.
- **Installed capacity** ($K_t^w$) from the Open Power System Data platform and ENTSO-E, used to normalize the feed-in profiles into capacity factors.

# 4. Results

We now present the empirical de-risking performance of the different contract designs. Using the simulated annual revenue streams described in Section 3, we first compare symmetric cash-flow risk across contracts using the coefficient of variation. We then examine the optimal leverage and implied WACC outcomes of our financial model, and analyze the resulting LCOE reductions. Finally, we decompose the contribution of each contract into subsidy and de-risking components.

## *4.1. Main results*

Figure 1 summarizes cash-flow risk across wind parks under each contract specification, measured by the coefficient of variation of annual revenues. First, wind parks operating under the merchant baseline exhibit very high CoVs, confirming that merchant exposure alone provides insufficient risk management for project finance. Second, all CfD designs substantially reduce cash-flow risk relative to the merchant case, but they do so to different degrees: two-sided and financial CfDs consistently deliver the lowest CoV values, while one-sided CfDs leave significantly more risk with investors. Third, two-sided CfDs not only achieve very low CoVs across specifications but also exhibit narrow CoV distributions, whereas the lowest overall cash-flow risk is achieved by park-optimal financial CfDs, even though these are optimized with respect to downside risk rather than CoV.

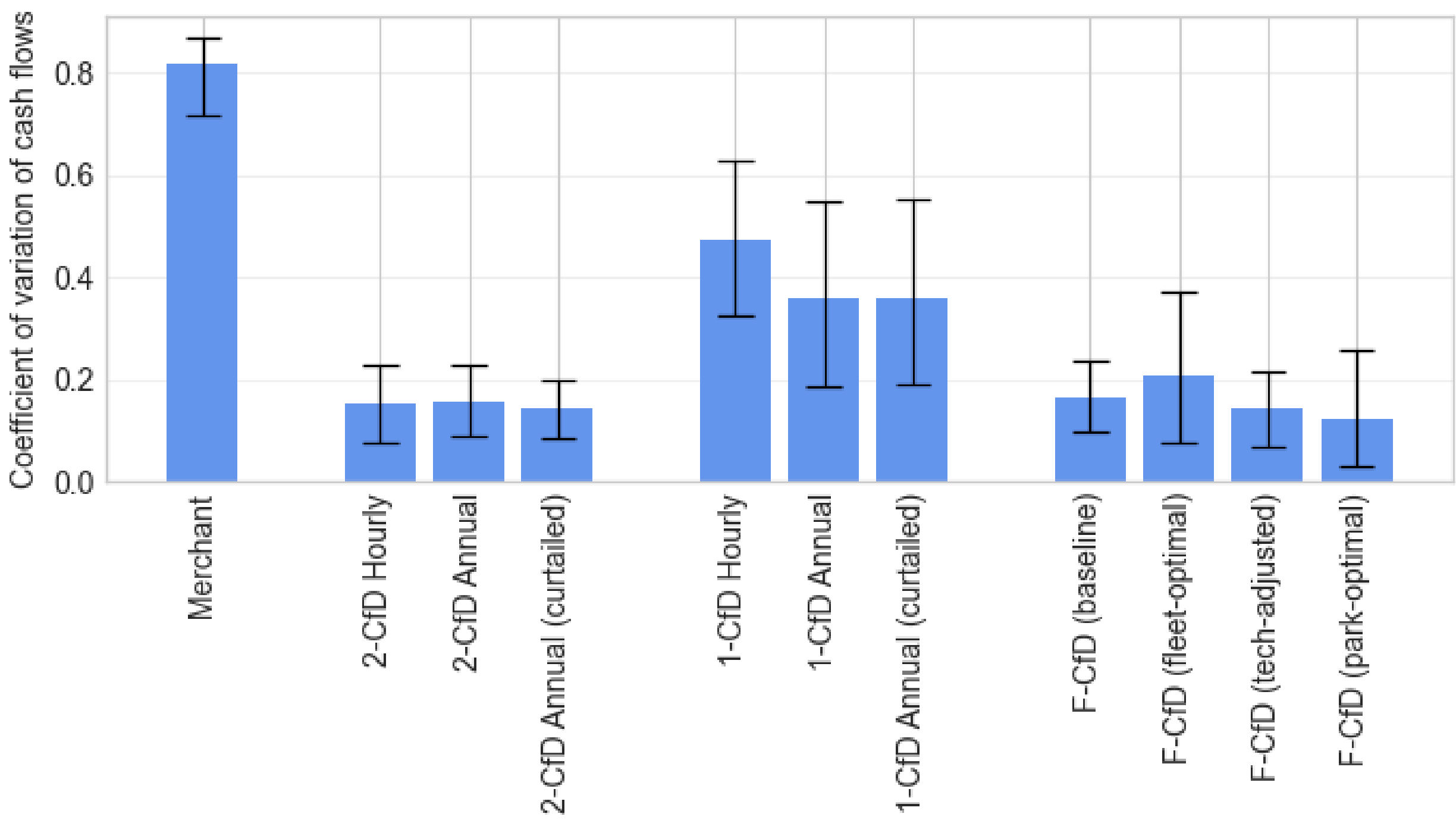

Figure 1. Average coefficient of variation of annual revenues across contract specifications. Bars show the average CoV for a given contract, and whiskers indicate the central 80% (10–90%) of the distribution.

A second insight from Figure 1 is that using an annual reference price and curtailing generation during negative-price hours *reduces* cash-flow risk for both two-sided and one-sided CfDs compared to hourly referencing, despite the introduction of some basis risk. This is the case if the covariance between the basic CfD payment ($q_{i,t}S$) and the annual adjustment ($q_{i,t}\left(S - p_t^{ref}\right)$) is negative. Though the negative covariance could be due to some properties of the assets in our sample and we do not model all the incentive implications of these contracts, using annual wind capture prices appears preferable, as it leads to both better incentives and lower risk for investors.

Feeding cash-flow patterns into our financial model, we calculate the equilibrium leverage ratios and WACC under each contract specification (Figure 2). Under the merchant baseline, most parks in our sample cannot afford any debt because at least one year's revenues would be insufficient to cover any debt service. This stark result reflects the mechanics of our financial model, which links the minimum annual cash flow to risk-free leverage, the significant land rents that need to be paid, and the fact that there are several years in our sample with particularly low electricity prices, including the 2020 Covid pandemic. For all CfD contracts, the outcome rates are consistent with the observed leverage ranges in real-world investments reported by Đukan et al. (2019) and Roth et al. (2021), who indicated that renewable energy projects under long-term contracts usually have debt sizes between 40% and 90% of the overall capital investment. Another finding consistent with Đukan et al. (2019) is that two-sided CfDs tend to lower the cost of capital more than one-sided CfDs. Regarding financial CfDs, specifications with endogenous contract factors have the highest leverage rates, whereas those that set the contract factor exogenously perform worse than two-sided CfDs.

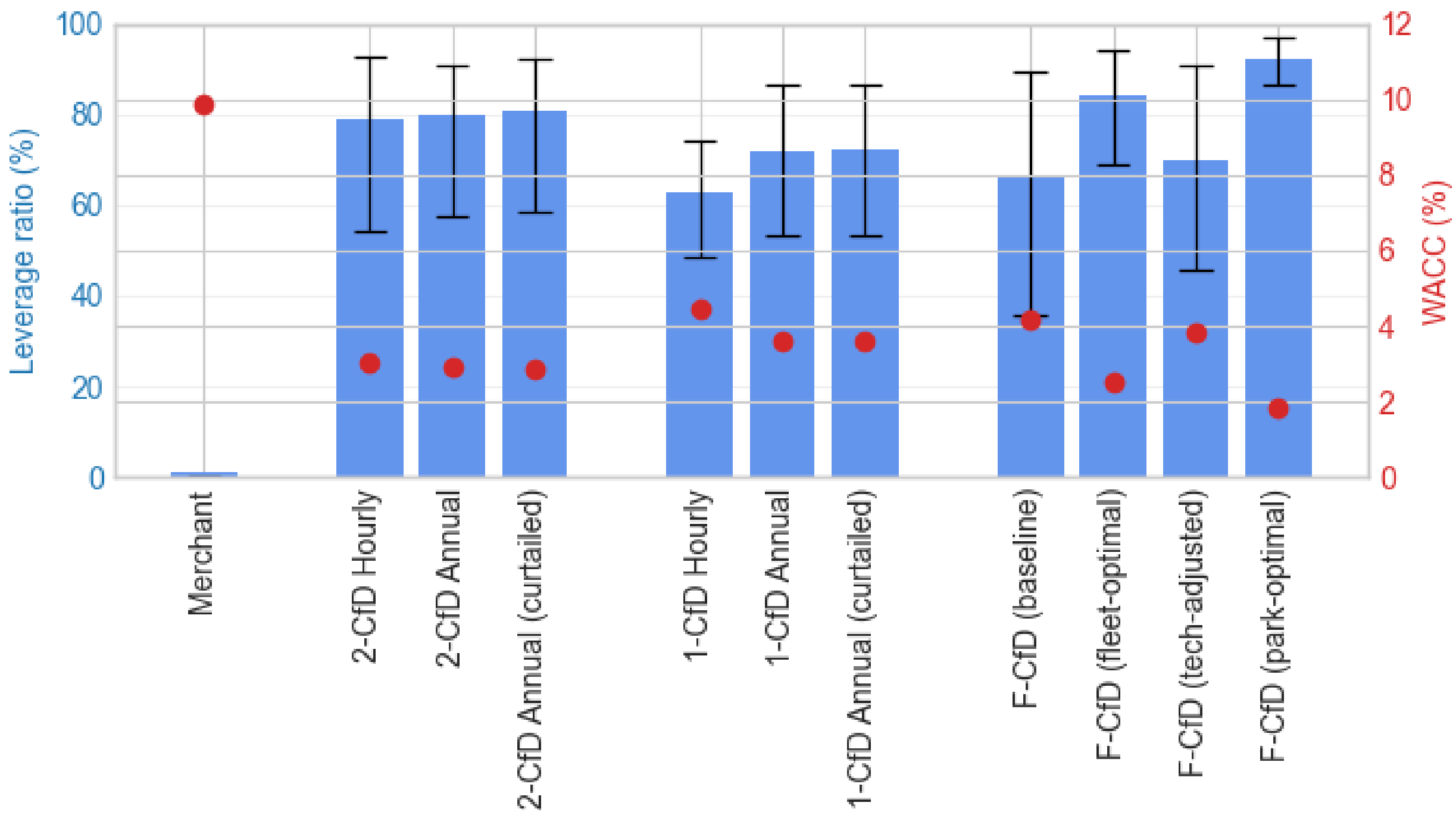

Figure 2. Financial indicators across wind parks under each contract type. Blue bars show the average leverage ratio for a given contract, and whiskers indicate the central 80% (10–90%) of the distribution. Red dots show the average WACC implied by the leverage ratio, based on the cost assumptions in Table 1.

From the WACC we calculate implied LCOE under each contract specification (Figure 3). All CfDs reduce LCOE relative to the merchant case by lowering financing costs, but the magnitude differs markedly across designs. Average LCOE reductions range between 40 and 70 €/MWh, based on our parameter assumptions about costs presented in Table 1. This means that between 30 to 40% of LCOE can be reduced by the better financing conditions affordable with the cash flows stemming from the contracts. Two-sided CfDs deliver the largest LCOE reductions among the classical designs, reflecting the complete price risk removal. One-sided CfDs also reduce LCOE relative to the merchant case, but less strongly, consistent with their weaker risk mitigation. As before, financial CfDs perform similar to one-sided CfDs or two-sided CfDs, depending on the contract factors. Optimal contract factors yield the lowest LCOEs of all contracts.

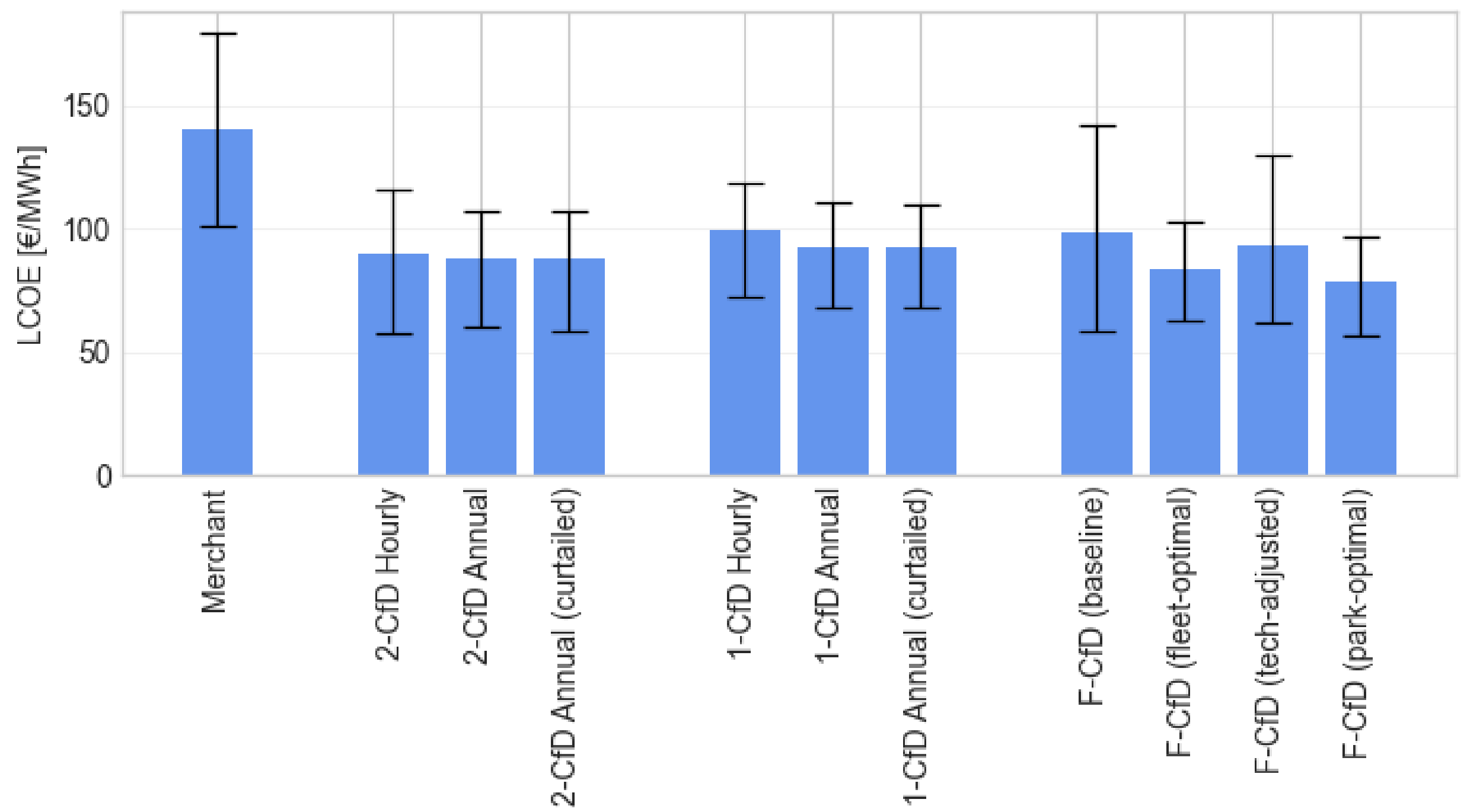

Figure 3. LCOE levels across wind parks under each contract type. Blue bars show the average leverage ratio for a given contract, and whiskers indicate the central 80% (10–90%) of the distribution.

Taken together, these results show that having any form of long-term public support plays a major role in accessing funds for investing in renewable energy, whereas the form of those contracts plays a second-order role. Still, differences in the range of 20 €/MWh are very significant. Hedging revenues through appropriately designed CfDs substantially reduces the LCOE relative to merchant exposure alone, and financial CfDs with well-chosen parameters can match or outperform two-sided CfDs in terms of risk reduction and financing conditions.

### *4.2. Contract effect decomposition*

As discussed in the introduction, long-term public contracts simultaneously subsidize revenues and hedge investors against cash-flow risk. We disentangle these two mechanisms by decomposing how each contract closes the "merchant gap", defined as the difference between the LCOE and the capture price under merchant exposure. For example, consider a project with a capture price of 30 €/MWh and an LCOE of 90 €/MWh; the merchant gap is 60 €/MWh. If a long-term contract reduced costs to 50 €/MWh through enabling low-cost financing, 40 €/MWh of the gap would be closed via risk reduction and remaining 20 €/MWh must come from outright subsidies. In this example, two thirds of the support effect arise from de-risking. This is a great deal for public finance, because only the subsidy strains public budgets.

The results are presented in Figure 4. The merchant gap that needs to be closed for a project to be viable is the same regardless of the contract, and it is on average 87 €/MWh, but we find that contracts differ in the relative size of their hedge and subsidy components. For the two-sided CfDs, the de-risking amounts to around 52 €/MWh, while the reminder, around 35 €/MWh, attributable to higher effective prices via the subsidy. For the fleet-optimal financial CfD, the hedging effect makes up 57 €/MWh, and for the park-optimal financial CfD, it reaches around 62 €/MWh, with 25 €/MWh of subsidy. This implies that for every 1€ spent in subsidizing parks under such contracts, societal returns are close to 3.5€.[3] These approximate figures underscore that contract design can strongly leverage public funds by targeting cash-flow risk rather than only price levels.

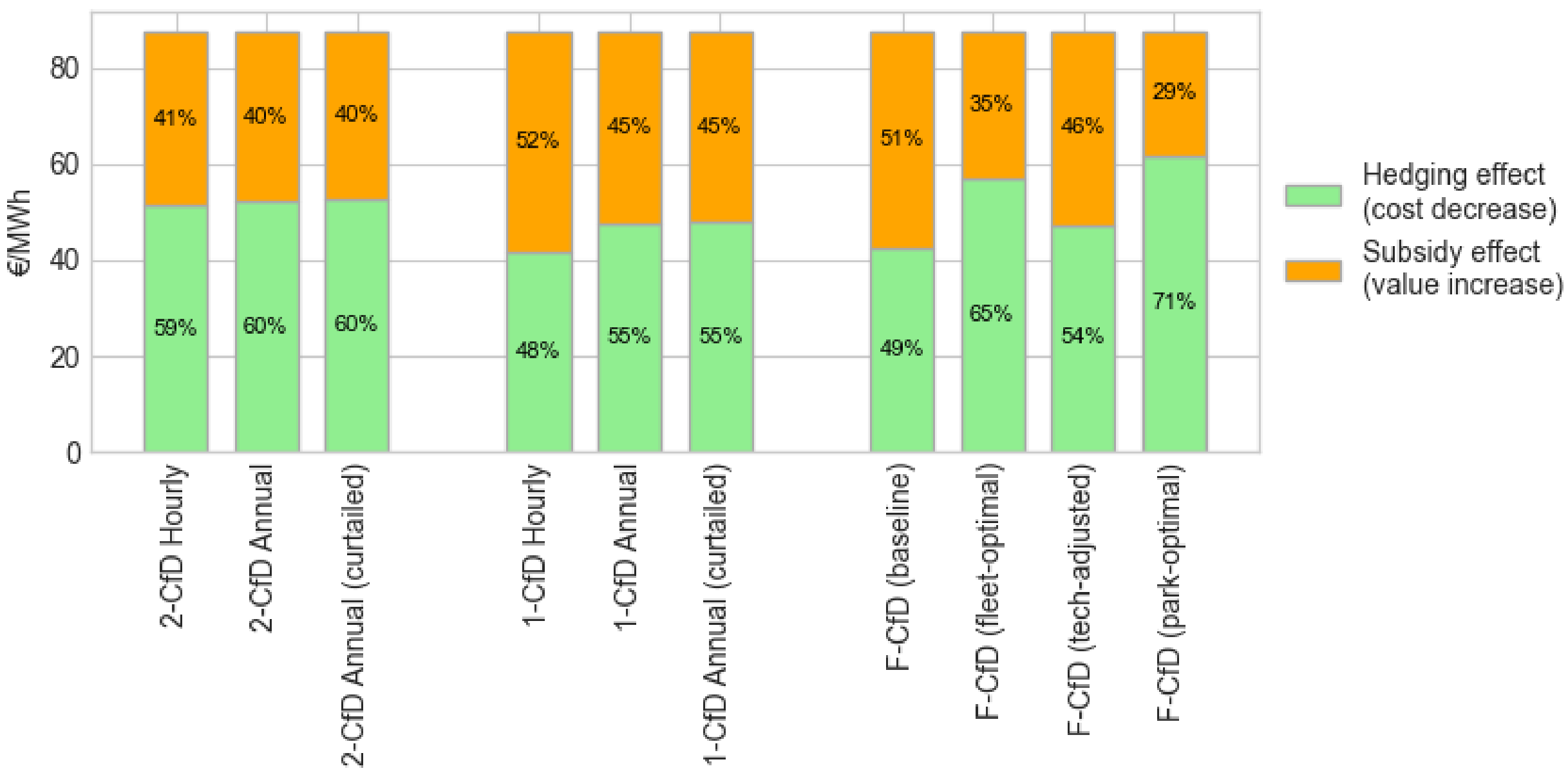

Figure 4. Bar plot showing the decomposition of the contract design closing the merchant gap into a subsidy effect (an increase in the price at which investors are selling the generated electricity on average) and a risk reduction effect (a decrease in LCOE due to better financing conditions).

### 4.3. Limitations

A first structural limitation is that, because we rely on historical data rather than a full market model, we do not capture the feedback from contract-specific dispatch incentives to market prices. During most of our sample period, German onshore wind generators were remunerated through feed-in tariffs or feed-in premiums, which encouraged production even at strongly negative spot prices. Under schemes with stronger price exposure (annual reference prices or financial CfDs) some of this output would likely have been withheld, raising prices in those hours and potentially changing contract performance. Our results should therefore be interpreted as partial-equilibrium revenue outcomes for given price paths.

A second limitation concerns our cost and financing assumptions. We apply homogeneous investment and operating costs and a common cost of debt and equity to all projects, even though our dataset spans different vintages, sites, and contractual arrangements. This implies that some parks will have LCOEs that are too high or too low relative to their true cost structure. This issue affects all results, from the fact that most merchant projects cannot afford any leverage and the size of the merchant gap to average LCOE levels. Because we cannot observe costs in a case-by-case basis, we use this homogeneous approach in general and explore parameter uncertainty in sensitivity analyses (see Appendix B).

[3] Note that this is only a first order approximation to the value of public funds spent in subsidizing wind energy. To accurately represent the marginal value of public funds as commonly done in the literature, one should account for additional factors such as the value of saved emissions and the costs of grid expansions or constraints induced by newly established renewable energy sources, which would vary based on the locations chosen for these investments.

Third, our focus is on project-level cash-flow risk and its implications for debt capacity, ignoring the impacts on equity returns and the attractiveness of investment under each contract. Higher leverage generally increases the tax shield and can raise equity IRRs, but also heightens exposure to financial distress, as emphasized by static trade-off theory. But in our setup, debt is risk-free by construction and there is no endogenous default. A natural extension would be to move from a deterministic DSCR rule to a stochastic framework in which debt is set such that the probability of violating a minimum DSCR remains below a threshold (e.g. a P90 test), but this would imply simulating revenue distributions instead of relying purely on historical data.

Regarding the empirical implementation of the financial CfD, we use a single nationwide reference generation profile. Alternative benchmarks based on location-specific wind models (Newbery, 2023; Schlecht et al., 2024) would alter the contract's basis risk, but not the core requirement that benchmarks remain sufficiently exogenous to preserve dispatch incentives.

Finally, we treat each wind park as a stand-alone investment. In practice, investors often hold portfolios of parks and technologies, so diversification may dampen project-specific volatility and could affect the ranking of contract designs through portfolio effects. Extending the analysis to diversified portfolios and alternative benchmark definitions is therefore a promising avenue for future work but lies beyond the scope of this paper.

# 5. Conclusion

This paper evaluates how different long-term public contracts mitigate investment risk for onshore wind projects, using real-world hourly production data from 63 wind parks over a ten-year period. We simulate project revenues under three stylized contract types –two-sided CfDs, one-sided CfDs, and financial CfDs– and quantify cash-flow risk using both the coefficient of variation of annual revenues and a stylized project-finance model focused on downside risk. In the spirit of structural capital-structure models and project-finance practice, debt capacity in our framework is determined by low-revenue states, with debt service required to be covered in all simulated years and equity bearing the residual risk. Under competitive auctions, contract levels are set so that both debt and equity participation constraints bind, yielding an implied LCOE for each contract design.

We find that financial CfDs consistently perform as well as, if not better, than one-sided and two-sided CfDs in reducing cash-flow risk and lowering LCOE, while preserving more exposure to market signals. When contract parameters are calibrated appropriately, these improvements do not require higher public spending. Financial CfDs thus emerge as a promising instrument that can relax the classic trade-off between risk mitigation and market integration: they substantially reduce investor risk without fully insulating producers from market dynamics.

Our decomposition of the merchant gap shows that a large share of the value created by these contracts can come from de-risking rather than from pure price support: for the most effective designs in our sample, roughly two thirds to three quarters of the gap is closed through lower financing costs, with the remainder due to higher effective prices. This suggests that, given the capital intensity of renewable energy investments, carefully designed hedging contracts can deliver substantially cheaper clean electricity for a given level of public expenditure, without necessarily implying excessive returns for generators. This is especially if support is allocated through competitive auctions, where lower risk is often associated with lower equity returns (Đukan et al., 2019)

For policymakers, these findings indicate that contract design, rather than more public spending in the form of higher subsidy levels, can simultaneously improve investment conditions and support the market integration of renewable energy sources. As renewable penetration continues to increase, aligning risk mitigation with efficient dispatch will be critical to sustain the energy transition while maintaining power systems and electricity markets operational. However, since financial CfDs reward price-adjusted generation rather than output alone, transitions from traditional production-based contracts (two- and one-sided CfDs) will create winners and losers depending on the production profiles of the assets within the current fleet of renewable generators.

# CRediT authorship contribution statement

Jorge Sánchez Canales: Writing – Original draft, Methodology, Visualization, Software, Project administration, Formal analysis, Conceptualization.

Lion Hirth: Writing – Review & editing, Funding acquisition, Conceptualization.

# Data and code availability

The data that support the findings of this study were provided by Enertrag SE, a renewable energy operator in Germany, under a data-sharing agreement and is not publicly available. The authors cannot share the actual due to contractual and confidentiality obligations.

The code used for data cleaning, analysis, and figure generation is publicly available in Github at jscanales/empirical_cfds.

# Declaration of use of generative AI and AI-assisted technologies

During the preparation of this work the authors used ChatGPT and DeepL Write to assist with writing the code used for the analysis, and to revise the writing of the original manuscript. After using these tools, the authors reviewed and edited the content as needed and take full responsibility for the content of the published article.

# Acknowledgements

This work is supported by the German Federal Ministry of Research, Technology and Space (BMFTR) via the Kopernikus project ARIADNE II (FKZ 03SFK5K0-2) and is part of the German Recovery and Resilience Plan (DARP), financed by NextGenerationEU, the European Union's Recovery and Resilience Facility (ARF).

The authors are grateful to Enertrag for generously providing access to the proprietary data used in this study. The views expressed in this paper are solely those of the authors and do not necessarily reflect those of the data provider.

# Appendix A. Data description

The original operational dataset contains 381 turbines across 93 wind parks, while the final estimation sample includes 268 turbines across 63 parks after applying the data-quality and balancing restrictions described in Section 3.4. Although our dataset provides uniquely detailed operational records, all assets are operated by a single company and are geographically concentrated in northeastern Germany. Figure A1 shows the location and commission year of the 63 wind parks in our final sample, highlighting both regional clustering and heterogeneity in turbine vintages.

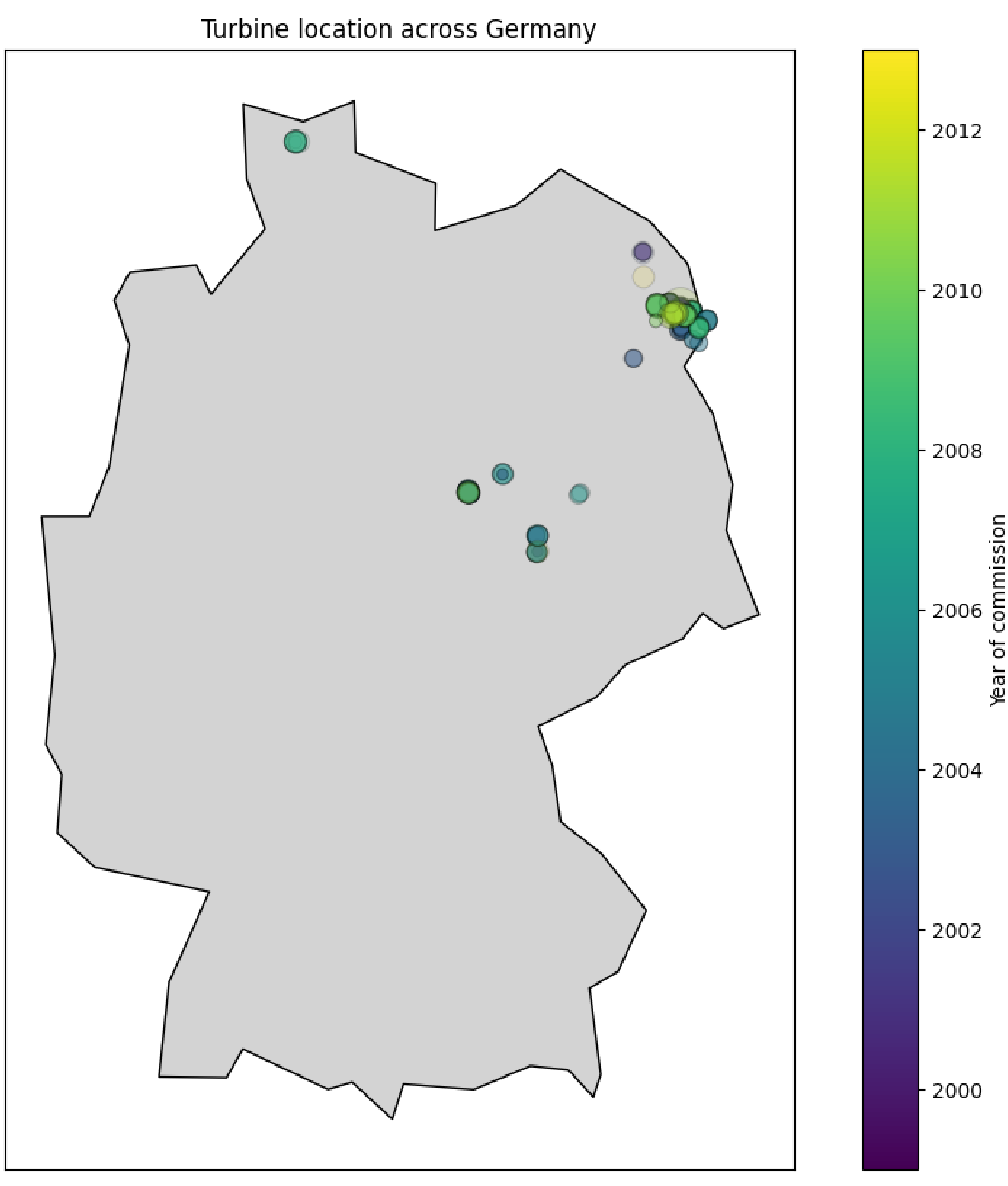

Figure A1. Location and commissioning year (color) and installed capacity (marker size) of wind parks in the sample.

To assess representativeness, we compare each park's hourly normalized production profile to the national onshore wind profile. Figure A2 reports the distribution of correlations between park-level and national output. Correlations range from about 0.3 to 0.8, with a mean around 0.6, suggesting substantial diversity in production patterns within our sample. In particular, the sample includes both "high-value" parks that earn more under market exposure than under a "neutral" fixed strike price equal to average prices and "high-volume" parks that perform better under such fixed remuneration than on the market. We regard this heterogeneity as representative of real-world investment opportunities.

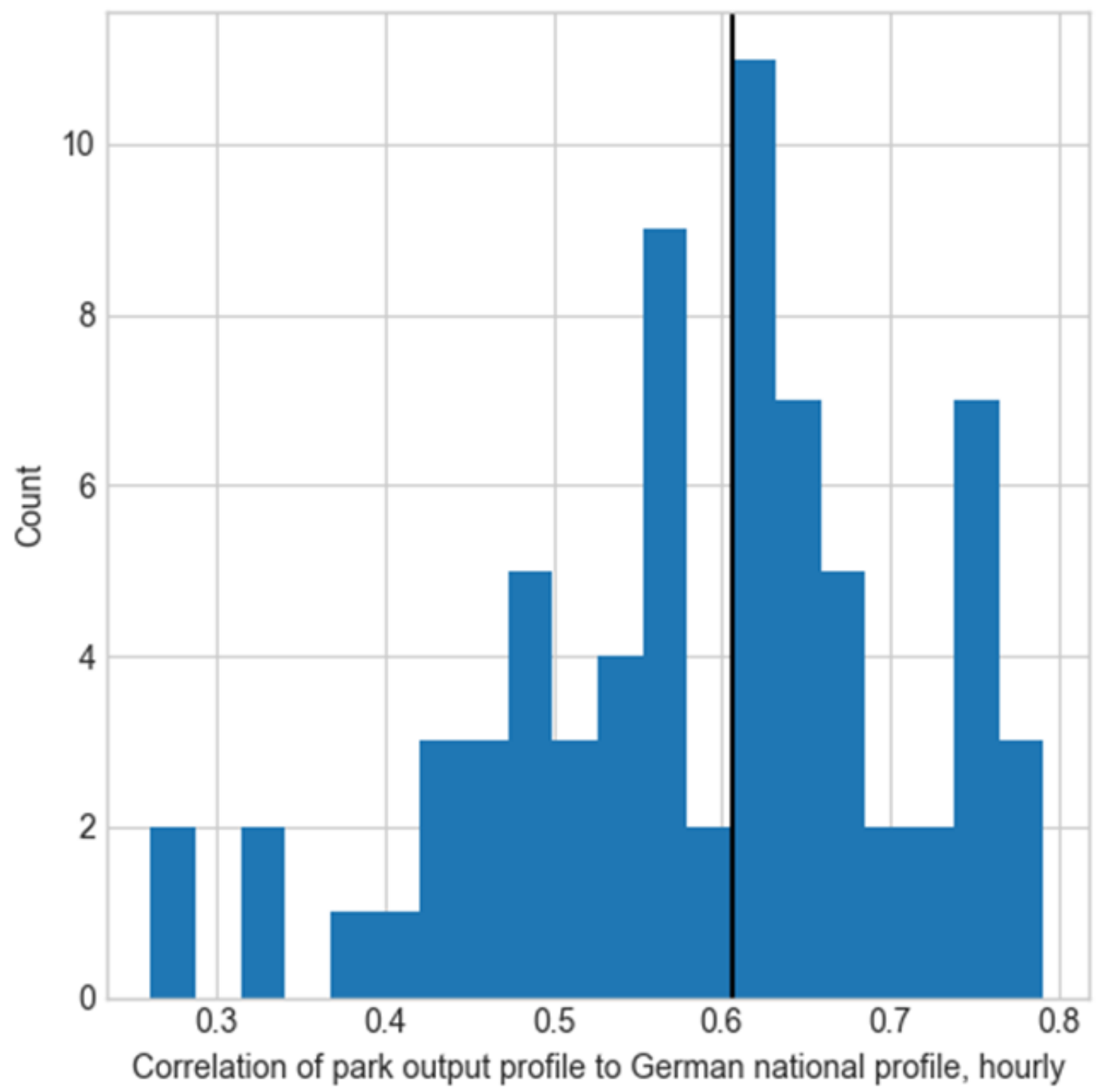

Figure A2. Histogram of correlations between each park's hourly normalized output and the national onshore wind profile; vertical line indicates the sample mean.

For each turbine we observe hourly power infeed, on-site wind speed, and a log of outages indicating cause and duration (e.g. grid constraints, maintenance, permitting, marketing decisions), including both full and partial curtailment of the potential output during the affected time window. Figure A3 summarizes the distribution of lost generation by outage category across turbines. On average, turbines in our sample produce around 91% of their potential output, with roughly 9% curtailed due to operational or grid-related issues, and grid-induced curtailment accounting for the largest share.

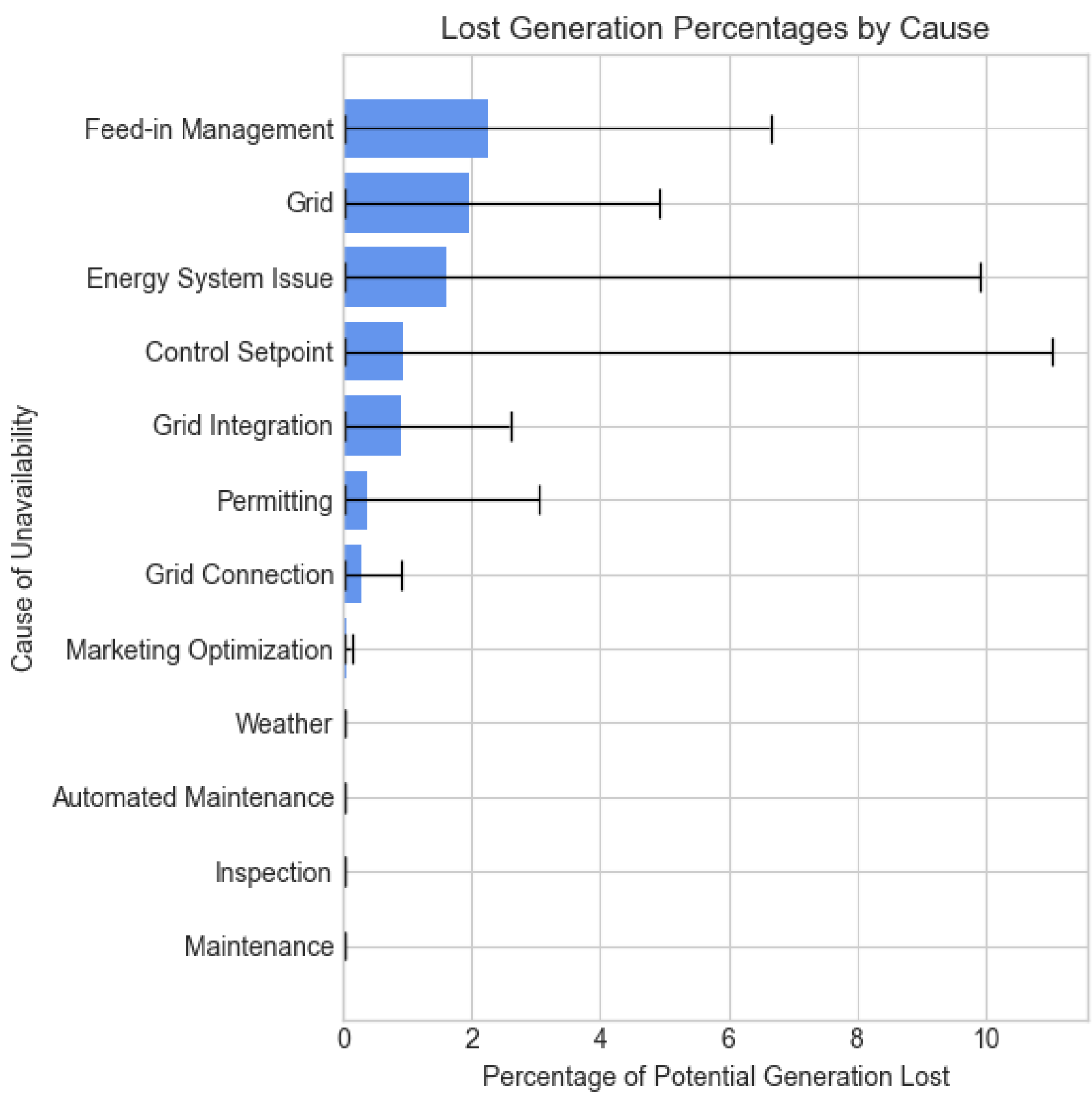

Figure A3. Distribution of lost generation (as percentage of potential output) by outage category across turbines in the sample. Bars show the average curtailed generation for a given cause, and whiskers indicate the central 90% (5–95%) of the distribution.

In Germany, energy curtailed for grid reasons is typically compensated as if it had been delivered, since these constraints are not under the control of investors. This makes it crucial to reconstruct potential (counterfactual non-curtailed) production when assessing the de-risking potential of different contract models. To ensure comparability across wind parks that are unequally affected by grid curtailment, and to align with the compensation regime, all simulated revenue streams in our analysis are based on potential generation, that is, the energy that would have been produced in the absence of grid-related curtailment, but keeping the other reasons for curtailment.

# Appendix B. Robustness

One limitation of our results concerns our treatment of investment and operating costs in the financial model. We have operationalized wind project investment and operating costs (IRENA, 2024) and WACC inputs (Đukan et al., 2025) using current levels and assuming homogeneous costs across all projects in the sample, even though our dataset covers investments over almost two decades, with different turbine vintages, site qualities, and contractual arrangements. In Germany, in particular, land leases (the major component of O&M costs) are often linked to realized revenues rather than being fixed ex ante, which would require more granular cost modeling. Because we cannot map project-level cost structures to each individual turbine or park, we opted for a common cost benchmark. We think this choice trades off realism for transparency and comparability across contract designs, but we address this limitation by conducting a sensitivity analysis. Figure B1 shows the results across a larger parameter space for investment and O&M costs. We show only a selection of contracts, those which perform best of each design.

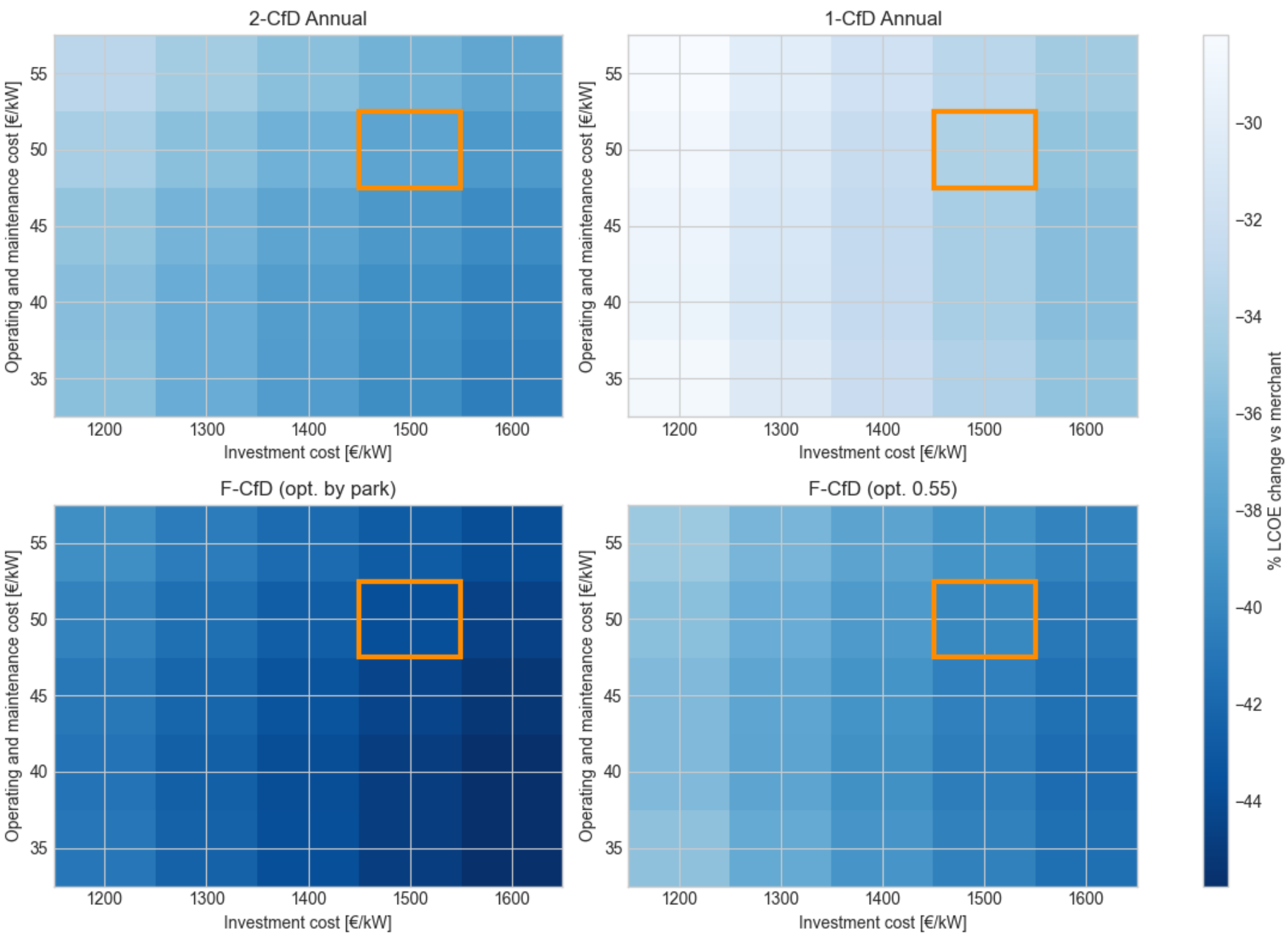

Figure B1. Heatmap showing the average reduction in LCOE across wind park as a function of the assumed investment costs $C^{fix}$ and operating and maintenance costs $C^{om}$. The grid square highlighted in orange reflects the baseline assumptions.

As we can see from Figure B1, the ordering of the cost reduction effect across contracts stays roughly the same across the parameter space. Two-sided and fleet-optimal financial CfDs lead to very similar reductions across cost assumptions, with the park-optimal financial CfD dominating in all grid points. The reduction of costs seems to be highest at high investment costs and low operating costs, as we would expect, and lowest in the opposite corner (low investment costs, high operating costs). In general, we learn that while the exact numbers of the risk reduction might be uninformative of the actual effects in the real world, the ordering of contracts is roughly the same across a reasonable parameter space.